\documentclass[twocolumn,aps,prl,amsmath,amssymb]{revtex4-1}
\usepackage[latin9]{inputenc}
\usepackage{amsmath}
\usepackage{amssymb}
\usepackage{graphicx}
\usepackage{wasysym}
\usepackage{esint}

\makeatletter
\usepackage{amssymb}
\usepackage{epsfig}
\usepackage{bm}
\usepackage{dcolumn}
\usepackage{color}

\makeatother

\begin{document}

\title{Low-momentum dynamic structure factor of a strongly interacting Fermi
gas at finite temperature: The Goldstone phonon and its Landau damping}

\author{Peng Zou$^{1}$, Hui Hu$^{2}$, and Xia-Ji Liu$^{2}$}

\affiliation{$^{1}$College of Physics, Qingdao University, Qingdao 266071, China}

\affiliation{$^{2}$Centre for Quantum and Optical Science, Swinburne University
of Technology, Melbourne, Victoria 3122, Australia}

\date{\today}
\begin{abstract}
We develop a microscopic theory of dynamic structure factor to describe
the Bogoliubov-Anderson-Goldstone phonon mode and its damping rate
in a strongly interacting Fermi gas at finite temperature. It is based
on a density functional approach - the so-called superfluid local
density approximation. The accuracy of the theory is quantitatively
examined by comparing the theoretical predictions with the recent
experimental measurements for the local dynamic structure factor of
a nearly homogeneous unitary Fermi gas at low transferred momentum
{[}S. Hoinka \textit{et al.}, Nat. Phys. \textbf{13}, 943 (2017){]},
without any free parameters. We calculate the dynamic structure factor
as functions of temperature and transferred momentum, and determine
the temperature evolution of the phonon damping rate, by considering
the dominant decay process of the phonon mode via scatterings off
fermionic quasiparticles. These predictions can be confronted with
future Bragg scattering experiments on a unitary Fermi gas near the
superfluid transition.
\end{abstract}


\maketitle
\textit{Introduction}. \textemdash{} The understanding of the density
fluctuation spectrum of superfluid $^{4}$He plays a central role
in the early development of quantum many-body physics \cite{Nozieres1990,Griffin1993}.
The modern concept of quasiparticles began with Landau's original
theory of superfluid $^{4}$He \cite{Khalatnikov1965}. The extensive
studies of either Brillouin and Raman light scattering \cite{Woods1973}
or inelastic neutron scattering \cite{Griffin1993} in such systems
lead to the discovery of phonons and rotons. In particular, the measurements
of the sound attenuation reveal the underlying decay mechanism of
quasiparticles. At low temperature (i.e., $T\lesssim0.6\textrm{ K}\sim0.3T_{c}^{\textrm{He}}$)
in the collisionless regime, the decay rate of phonons $\Gamma$ is
due to the three-phonon Landau-Beliaev process and exhibits a characteristic
$\omega T^{4}$ dependence on the phonon frequency $\omega$ \cite{Woods1973}.
At higher temperature ($T\gtrsim1.0\textrm{ K}\sim0.5T_{c}^{\textrm{He}}$),
superfluid $^{4}$He crosses over to the hydrodynamic regime and the
damping rate of phonons instead shows a quadratic $\omega^{2}$ dependence
\cite{Khalatnikov1965}.

After nearly fifty years, the community of quantum physics welcomes
the arrival of another strongly interacting many-body system \cite{OHara2002},
a unitary Fermi gas at the cusp of the crossover from Bose-Einstein
condensates (BEC) to Bardeen-Cooper-Schrieffer (BCS) superfluids \cite{Randeria2014}.
This novel fermionic superfluid is unique, owing to the unprecedented
accuracy in tuning almost all the controlling parameters of the system
\cite{Bloch2008}. To date, there are already a number of milestone
observations of a unitary Fermi gas, confirming its high-temperature
superfluidity \cite{Regal2004}, measuring the zero-temperature equation
of state \cite{Navon2010}, revealing the universal thermodynamics
\cite{Ho2004,Hu2007,Nascimbene2010,Ku2012} and probing the second
sound \cite{Sidorenkov2013}. The density fluctuation spectrum is
also measured, however, restricted to collective oscillations with
\emph{discrete} frequencies \cite{Kinast2004,Bartenstein2004,Hu2004,Altmeyer2007},
due to the very existence of a harmonic trapping potential that is
necessary to hold atoms from escaping. Only most recently, the density
excitation spectrum of a nearly \emph{homogeneous} unitary Fermi gas
has been obtained at Swinburne University of Technology, by applying
the low-momentum two-photon Bragg spectroscopy to determine the local
dynamic structure factor near the trap center \cite{Hoinka2017}.
The purpose of this work is to present a microscopic theory that \emph{quantitatively}
explains the observed Bogoliubov-Anderson-Goldstone phonon mode and
to provide reliable theoretical predictions on the phonon damping
for future experimental confirmation.

The development of a quantitative description of the density response
of a strongly interacting Fermi superfluid is by no means an easy
task \cite{Minguzzi2001,Combescot2006,Stringari2009,Zou2010,Guo2010,Palestini2012,Hu2012,He2016,Zou2016,Vitali2017}.
There is no small parameter to control the precision of the theory
due to the divergent scattering length $a_{s}$ in the unitary limit
\cite{Randeria2014}. For the experimental work at Swinburne, the
data have been qualitatively understood using a standard random-phase-approximation
(RPA) theory, with a modified chemical potential as a fitting parameter
\cite{Hoinka2017}. The calculated spectrum overestimates the phonon
peak (i.e., more than twice in height) and accounts for only two-thirds
the measured width. This good but somewhat unsatisfactory agreement
is partly because of the violation in the $f$-sum rule, as a result
of the large modification to the mean-field chemical potential \cite{Hoinka2017}.
The quantitative agreement between our microscopic theory and experiment
without any adjustable parameters, as found in this work, is therefore
highly nontrivial. 

The establishment of an accurate density response theory also allows
us to identify the main decay mechanism for Goldstone phonons. In
contrast to the three-phonon Landau-Beliaev processes as previously
suggested \cite{Kurkjian2017AOP}, we clarify that the phonon damping
is dominated by the inelastic process of absorption or emission by
fermionic quasiparticles \cite{Zhang2011,Castin2017,Kurkjian2017}.
We determine the inverse quality factor $\Gamma/\omega$ of a unitary
Fermi gas as functions of temperature and transferred momentum. These
predictions could be readily examined in state-of-art experiments
in cold-atom laboratories.

\textit{Density response theory within SLDA}. \textemdash{} We start
by briefly reviewing the superfluid local density approximation (SLDA)
of a unitary Fermi gas \cite{Yu2003,Bulgac2007} and the resulting
improved SLDA-RPA theory for density response functions \cite{Zou2016}.
As the $s$-wave scattering length $a_{s}$ diverges in the unitary
limit, the low-energy physics of the system can be well-governed by
a regularized energy density functional $\mathcal{E}[\tau_{c}(\mathbf{r},t),n(\mathbf{r},t),\nu_{c}(\mathbf{r},t)]$,
\begin{equation}
\mathcal{E}\left[\tau_{c},n,\nu_{c}\right]=\frac{\tau_{c}}{2m}+\beta\frac{3\left(3\pi^{2}\right)^{2/3}}{10m}n^{5/3}+g_{\textrm{eff}}\left|\nu_{c}\right|^{2},
\end{equation}
where $\tau_{c}=2\sum_{\mathbf{\left|k\right|}<\Lambda}\left|\nabla v_{\mathbf{k}}\right|^{2}$
is the kinetic density, $n=2\sum_{\mathbf{\left|k\right|<\Lambda}}\left|v_{\mathbf{k}}\right|^{2}$
the number density, $\nu_{c}=\sum_{\mathbf{\left|k\right|<\Lambda}}u_{\mathbf{k}}v_{\mathbf{k}}^{*}$
the anomalous Cooper-pair density, $u_{\mathbf{k}}(\mathbf{r},t)$
and $v_{\mathbf{k}}(\mathbf{r},t)$ are the Bogoliubov quasiparticle
wavefunctions, to be determined by solving a generalized Bogoliubov-de
Gennes equation for momentum $\mathbf{k}$ below the cut-off momentum
$\Lambda$ \cite{Zou2016,Yu2003,Bulgac2007,CutOffMomentumNote} ,
and $g_{\textrm{eff}}^{-1}\equiv mn^{1/3}/\gamma-\sum_{\mathbf{\left|k\right|}<\Lambda}m/\mathbf{k}^{2}$
is the inverse effective coupling constant. The form of the above
energy density functional is motivated by the scale invariance that
is satisfied at unitarity and the two parameters $\beta$ and $\gamma$
can be uniquely fixed by requiring that the resulting chemical potential
$\mu$ and pairing gap $\Delta$ agree with those calculated by microscopic
theories or measured experimentally \cite{EffectiveMassNote}. As
the pairing gap is related to the anomalous density $\nu_{c}$ by
$\Delta(\mathbf{r},t)=-g_{\textrm{eff}}\nu_{c}(\mathbf{r},t)$, we
may rewrite the interaction part of the density functional as,
\begin{equation}
\mathcal{E}_{\textrm{int}}=\beta\frac{3\left(3\pi^{2}\right)^{2/3}}{10m}\left[n\left(\mathbf{r},t\right)\right]^{5/3}+\frac{\left|\Delta\left(\mathbf{r},t\right)\right|^{2}}{g_{\textrm{eff}}}.
\end{equation}
Let us consider the fluctuations in the number densities $n_{\uparrow}(\mathbf{r},t)$,
$n_{\downarrow}(\mathbf{r},t)$, and Cooper-pair density $\nu_{c}(\mathbf{r},t)$
and its complex conjugate $\nu_{c}^{*}(\mathbf{r},t)$, to be collectively
denoted as $\delta n_{i}$ or $\delta n_{j}$ ($i,j=\uparrow,\downarrow,c,c^{*}$).
Here, we have split the total density $n(\mathbf{r},t)$ into the
spin-up and spin-down components to allow the calculation of spin-density
dynamic structure factor. These local fluctuations induce a self-generated
mean-field potential $\sum_{j}E_{ij}^{I}\delta n_{j}$, where $E_{ij}^{I}=(\delta^{2}\mathcal{E}_{\textrm{int}}/\delta n_{i}\delta n_{j})$
\cite{Minguzzi2001,Stringari2009}. As a result, the dynamical response
function takes the standard RPA form, $\chi=\chi^{0}[1-\chi^{0}E^{I}]^{-1}$,
where $\chi^{0}$ is the bare response function without the inclusion
of the induced potential \cite{Minguzzi2001}. The density response
function is a summation of $\chi_{ij}$ in the density channel, i.e.,
$\chi_{nn}(\mathbf{k},\omega+i0^{+})=\chi_{11}+\chi_{12}+\chi_{21}+\chi_{22}=2(\chi_{11}+\chi_{12})$.
The dynamic structure factor is given by the imaginary part of $\chi_{nn}$,
i.e., $S\left(\mathbf{k},\omega\right)=-\textrm{Im}\chi_{nn}/[\pi(1-e^{-\hbar\omega/k_{B}T})]$.

In our previous work \cite{Zou2016}, we have derived the expression
for the matrix $E^{I}$ and calculated the dynamic and static structure
factor at \emph{zero} temperature. We have found that the SLDA-RPA
dynamic structure factor satisfies the important $f$-sum rule and
compressibility sum rule and the static structure factor $S(k)$ agrees
very well with the latest quantum Monte Carlo result for $k<k_{F}$
\cite{Carlson2014}, where $k_{F}$ is the Fermi wavevector. The excellent
agreement strongly indicates that our SLDA-RPA theory could be quantitatively
reliable near the unitary limit at low temperature. Here, we confirm
this anticipation by the more stringent comparison with the recent
experimental measurements at \emph{finite} but low temperature \cite{Hoinka2017},
without any free parameters.

\begin{figure}
\centering{}\includegraphics[width=0.48\textwidth]{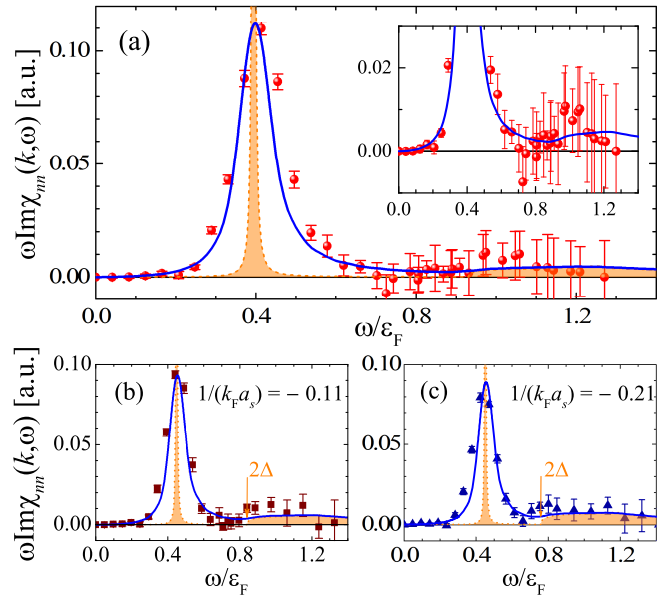}\caption{\label{fig1_TheoryExperiment} (color online). The comparison between
the SLDA-RPA theoretical predictions and the experimental data for
$-\omega\textrm{Im}\chi_{nn}(k,\omega)/\pi\equiv\omega(1-e^{-\hbar\omega/k_{B}T})S(k,\omega)$
at three sets of experimental conditions \cite{Hoinka2017}: (a) the
unitary limit with $1/(k_{F}a_{s})=0$, $T=0.09T_{F}$ and $k=0.55k_{F}$,
(b) $1/(k_{F}a_{s})=-0.11$, $T=0.082T_{F}$ and $k=0.60k_{F}$, and
(c) $1/(k_{F}a_{s})=-0.21$, $T=0.078T_{F}$ and $k=0.59k_{F}$. The
blue solid lines take into account the spectral broadening due to
the finite duration of the Bragg pulse and due to the slight density
inhomogeneity around the trap center (see text for more details),
while the orange dotted lines with shadow report the original SLDA-RPA
results before the convolutions with the spectral broadening functions.
The inset in (a) highlights the comparison near the pair-breaking
excitations. In (b) and (c), the pair-breaking energy $2\Delta$ is
explicitly indicated by an arrow. The theoretical predictions are
normalized to have the same area as the experimental results, according
to the $f$-sum rule $-\int_{0}^{\infty}\omega\textrm{Im}\chi_{nn}(k,\omega)/\pi d\omega=k^{2}/(2m)$,
which should be satisfied both theoretically and experimentally. }
\end{figure}

\textit{Quantitative comparison}. \textemdash{} To foster the comparison,
it should be noted that the experimentally measured density fluctuation
spectrum includes \emph{instrumental} broadening due to the finite
duration of the Bragg pulse. It can be viewed as the intrinsic response
function convoluted with a sinc line shape \cite{Brunello2001,Blakie2002}:
\begin{equation}
\textrm{Im}\chi_{nn}^{(\textrm{Br})}\left(\mathbf{k},\omega\right)=\intop_{-\infty}^{\infty}d\omega'\frac{\textrm{Im}\chi_{nn}\left(\mathbf{k},\omega'\right)}{\pi\sigma_{B}}\textrm{sinc}^{2}\left[\frac{\omega-\omega'}{\sigma_{B}}\right],
\end{equation}
where $\textrm{sinc}(x)\equiv\sin(x)/x$ and the energy resolution
$\sigma_{B}=2/\tau_{\textrm{Br}}$ is set by the pulse duration ($\tau_{\textrm{Br}}\simeq1.2$
ms \cite{Hoinka2017}). In addition, the experimental spectrum also
includes broadening arising from the slight density inhomogeneity
near the trap center ($\delta n/n\sim0.08$) \cite{Hoinka2017}. As
the Fermi energy $\varepsilon_{F}\propto n^{2/3}$ at unitarity, we
estimate the variation $\delta\varepsilon_{F}\simeq0.06\varepsilon_{F}$.
We account for this trap-induced broadening by further convoluting
the spectrum with a Lorentzian line shape,
\begin{equation}
\textrm{Im}\chi_{nn}^{(\textrm{exp})}\left(\mathbf{k},\omega\right)=\intop_{-\infty}^{\infty}d\omega'\frac{\left(\sigma_{T}/\pi\right)\textrm{Im}\chi_{nn}^{(\textrm{Br})}\left(\mathbf{k},\omega'\right)}{\left[\left(\omega-\omega'\right)^{2}+\sigma_{T}^{2}\right]},
\end{equation}
where $2\sigma_{T}=\delta\varepsilon_{F}$ is the full width at half
maximum.

\begin{figure}[t]
\centering{}\includegraphics[width=0.45\textwidth]{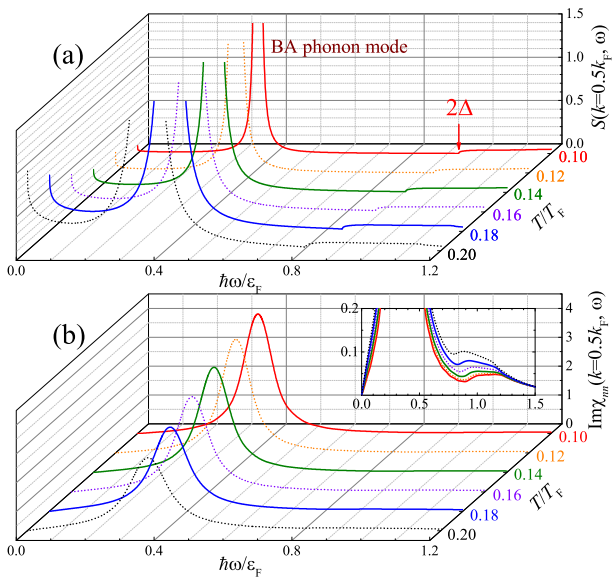}
\caption{\label{fig2_dsf050kf} (color online). Temperature evolution of the
dynamic structure factor of a unitary Fermi gas at the transferred
momentum $k=0.5k_{F}$: (a) the original SLDA-RPA results $S(k,\omega)$
and (b) the more experimentally relevant predictions $-\textrm{Im}\chi_{nn}(k,\omega)/\pi\equiv(1-e^{-\hbar\omega/k_{B}T})S(k,\omega)$,
after convolutions with the spectral broadening functions. The inset
in the lower panel highlights the density response near the pair-breaking
threshold $2\Delta$.}
\end{figure}

\begin{figure}[t]
\centering{}\includegraphics[width=0.4\textwidth]{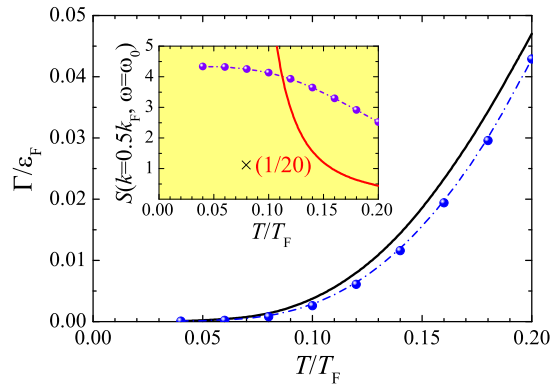}\caption{\label{fig3_DampingWidth} (color online) The damping width of the
Goldstone mode of a unitary Fermi gas at $k=0.5k_{F}$, as a function
of temperature. The black solid line shows the original SLDA-RPA results.
The blue dot-dashed line with circles reports $\delta\Gamma=\Gamma-\Gamma_{0}$,
where $\Gamma(T)$ is obtained by convoluting the SLDA-RPA dynamic
structure factor with the spectral broadening functions and $\Gamma_{0}=\Gamma(T\rightarrow0)\simeq0.1\epsilon_{F}$
is the background width of the spectral broadening. The inset shows
the dynamic structure factor at the peak position $\omega_{0}$, without
(red line) or with convolutions (purple dot-dashed line with circles).
For better illustration, the original SLDA-RPA result (red line) has
been reduced by a factor of $20$. }
\end{figure}

Fig. \ref{fig1_TheoryExperiment} presents the comparison between
the experimental spectra (symbols) and theoretical predictions obtained
after performing the two convolutions (solid lines) \cite{SpectrumConvolutionNote},
for three sets of parameters. We have used the zero-temperature equations
of state given by a Gaussian pair fluctuation theory \cite{Hu2006}
as the inputs to determine the parameters $\beta$ and $\gamma$ at
different interaction strengths \cite{Zou2016}. The results for density
response function before the convolutions are also shown by dashed
lines with shadow. There is an excellent agreement for the \emph{entire}
spectrum, both at the unitary limit (a) and near unitarity (b, c).
Our theory greatly improves the previous RPA explanation \cite{Hoinka2017},
in the sense that (i) it removes the necessity of introducing a fitting
parameter, to fit the measured peak position $\omega_{0}(k)$ of the
Goldstone phonon mode; (ii) it fully accounts for the observed width
and height of the peak; and (iii) it does not need to scale, in order
to match the amplitude of the measured pair-breaking excitations at
$\omega\sim2\Delta$. The last point is particularly clear in Fig.
\ref{fig1_TheoryExperiment}, as the broad single-particle excitations
are essentially unaffected by the instrumental broadening and trap
inhomogeneity. The agreement is therefore highly non-trivial and actually
it emphasizes the importance of the significant renormalization of
single-particle behavior, due to the strong pairing effect, which
is indeed taken into account in our theory via a density functional
approach.

\textit{T-dependence of the Goldstone mode}. \textemdash{} By establishing
the reliability of our SLDA-RPA theory, we turn to consider the temperature
evolution of the dynamic structure factor in the unitary limit, as
shown in Fig. \ref{fig2_dsf050kf}(a). The experimentally measurable
density response (after convolutions) is reported in Fig. \ref{fig2_dsf050kf}(b).
Here and in the following, we have taken the experimentally determined
chemical potential $\mu=0.376\varepsilon_{F}$ \cite{Ku2012} and
pairing gap $\Delta=0.47\varepsilon_{F}$ \cite{Hoinka2017} to fix
the parameters $\beta$ and $\gamma$. As temperature increases, it
is apparent that the phonon peak in the spectrum becomes wider and
lower, suggesting that the intrinsic width of the phonon mode becomes
significant. Moreover, the single-particle excitations at around $\omega=2\Delta$
are enhanced (see the inset in Fig. \ref{fig2_dsf050kf}(b)). Focusing
on the Goldstone mode, we present the temperature dependence of its
width and peak height in Fig. \ref{fig3_DampingWidth}. It is interesting
that, by subtracting a background width $\Gamma_{0}\sim0.1\varepsilon_{F}$
due to the instrumental and inhomogeneity broadenings, the width to
be experimentally measured (symbols with dot-dashed line) is roughly
equal to the \emph{intrinsic} width of the phonon peak in the dynamic
structure factor (solid line). This simply indicates that the intrinsic
width of the Goldstone mode could be directly read from the measured
width, with reasonable accuracy. 

\begin{figure}[t]
\centering{}\includegraphics[width=0.4\textwidth]{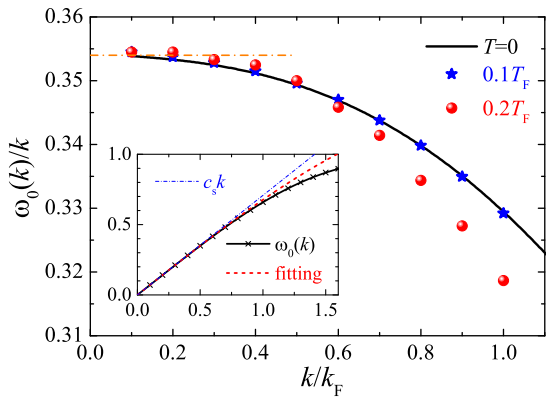}\caption{\label{fig4_DispersionRelation} (color online). The phonon phase
velocity $\omega_{0}(k)/k$ of a unitary Fermi gas at three different
temperatures as indicated. The horizontal orange dot-dashed line indicates
the sound velocity at zero temperature $c_{s}=(\xi/3)^{1/2}v_{F}\simeq0.354v_{F}$.
The inset presents the zero-temperature dispersion relation $\omega_{0}(k)$,
together with the leading order contribution $\omega_{0}(k)\simeq c_{s}k$
(blue dot-dashed line). The red dashed line is the fitting curve that
takes into account the next order, i.e., $\omega_{0}(k)\simeq c_{s}k(1+\zeta k^{2})$,
where $\zeta=-0.044(3)k_{F}^{-2}<0$.}
\end{figure}

\begin{figure}[t]
\centering{}\includegraphics[width=0.45\textwidth]{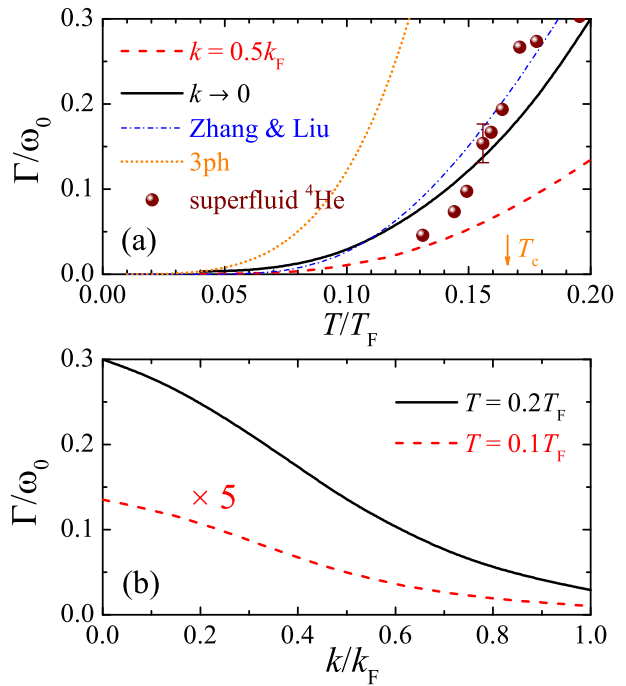}\caption{\label{fig5_InverseQualityFactor} (color online) (a) Temperature
dependence of the inverse quality factor $\Gamma/\omega_{0}$ of a
unitary Fermi gas at $k\rightarrow0$ (black solid line) and at $k=0.5k_{F}$
(red dashed line). For comparison, we show the result by Zhang and
Liu due to scatterings off fermionic quasiparticles (blue dot-dashed
line) \cite{Zhang2011,ZhangLiuResultNote} and the prediction Eq.
(\ref{eq:IQF3ph}) by Kurkjian, Castin and Sinatra due to the three-phonon
interaction process (orange dotted line) \cite{Kurkjian2017AOP},
both of which are applicable in the limit of small momentum $k\rightarrow0$.
We show also $\Gamma/\omega_{0}$ of superfluid $^{4}$He at $k=0.4\textrm{ }\textrm{Å}^{-1}$
and at saturated vapor pressure (circles) \cite{Stirling1990}, with
temperature rescaled as $T\rightarrow(T/T_{c}^{\textrm{He}})\times T_{c}$.
The arrow indicates the transition temperature of a unitary Fermi
gas $T_{c}\simeq0.167T_{F}$. (b) The inverse quality factor $\Gamma/\omega_{0}$
of a unitary Fermi gas at two temperatures $T=0.1T_{F}$ and $T=0.2T_{F}$,
as a function of the transferred momentum. The result at $T=0.1T_{F}$
has been amplified by a factor of $5$ for better visualization.}
\end{figure}

\textit{Landau damping}. \textemdash{} We now turn to discuss the
intrinsic width or damping rate of the Goldstone phonon mode in greater
detail. What is the main mechanism responsible for damping? Physically,
there are three possible sources that we may consider: (1) the three-phonon
Landau-Beliaev process $\phi\longleftrightarrow\phi\phi$, where $\phi$
is the annihilation field operator of phonons; (2) the four-phonon
Landau-Khalatnikov process $\phi\phi\longleftrightarrow\phi\phi$;
and (3) the inelastic process of absorption or emission by the single-particle
excitations. All these processes are responsible in the case of superfluid
$^{4}$He. For example, at low temperature the three-phonon process
is kinematically allowed by the anomalous dispersion of the phonon
mode (i.e., $\omega_{0}(k)=c_{s}k(1+\zeta k^{2})$ with a positive
$\zeta>0$) at $k<k_{c}\sim0.55\textrm{Å}^{-1}$. The much weaker
four-phonon process is possible at $k>k_{c}$. For temperatures above
$1\textrm{ K},$ the last inelastic process becomes favorable by scattering
from thermally excited rotons.

For a unitary Fermi gas, Kurkjian and co-workers suggested that the
three-phonon process is the dominant decay mechanism, since the standard
RPA theory predicts a positive $\zeta$ \cite{Kurkjian2016}, and
derived an elegant expression for the inverse quality factor at low
temperature \cite{Kurkjian2017AOP}:
\begin{equation}
\frac{\Gamma}{\omega_{0}}\overset{k\rightarrow0}{=}\frac{16\pi^{5}\sqrt{3}}{405}\xi^{3/2}\left(\frac{k_{B}T}{mc_{s}^{2}}\right)^{4}\simeq1.2\times10^{3}\left(\frac{T}{T_{F}}\right)^{4},\label{eq:IQF3ph}
\end{equation}
where we have used the Bertsch parameter $\xi=0.376$ \cite{Ku2012}.
This observation, however, is not conclusive, since our more accurate
SLDA-RPA theory gives a negative $\zeta$ at both zero and finite
temperatures, as shown in Fig. \ref{fig4_DispersionRelation}. Another
qualitative $\epsilon$-expansion theory provides a similar negative
$\zeta$ at unitarity \cite{Haussmann2009}. On the other hand, it
is known that the three-phonon process in $^{4}$He is only relevant
at low temperature (i.e., $T<1.0\textrm{ K}\sim0.5T_{c}^{\textrm{He}}$)
\cite{Griffin1993,Woods1973}. It is thus unlikely to be the main
damping source in a unitary Fermi gas for $T>0.5T_{c}\sim0.09T_{F}$.
From the above considerations, we would like to argue that the inelastic
scatterings of phonons from fermionic quasiparticles causes their
damping at the temperature region $T\apprge0.1T_{F}$, which is of
great experimental interest. 

If this is true, we expect that the damping rate $\Gamma$ will be
approximately proportional to the number of fermionic quasiparticles
present, i.e., $\Gamma\propto e^{-E_{\textrm{min}}/k_{B}T}$, where
$E_{\textrm{min}}\equiv\min\{E(k)\}$ is the minimum energy of single
particles that satisfies some momentum and energy conservation requirements
to allow inelastic scatterings. More quantitatively, an analytic expression
for $\Gamma$ was first derived by Zhang and Liu by considering the
phase fluctuations at small momentum $k\rightarrow0$ within RPA \cite{Zhang2011,ZhangLiuResultNote}.
Most recently, it was improved by Kurkjian \textit{et al}. by taking
into account the modified single-particle dispersion \cite{Castin2017}
and/or the amplitude fluctuations \cite{Kurkjian2017}. Our SLDA-RPA
theory may give a better prediction, starting from a more fundamental
microscopic approach, although we cannot obtain an analytic expression
for $\Gamma/\omega_{0}$. Our results at $k\rightarrow0$ and $k=0.5k_{F}$
are reported in Fig. \ref{fig5_InverseQualityFactor}(a) using solid
and dashed lines, respectively. The result at small momentum qualitatively
agrees with the earlier prediction by Zhang and Liu (dot-dashed line),
as one may anticipate. The damping rate decreases steadily with increasing
transferred momentum $k$ as predicted earlier \cite{Kurkjian2017},
as can be seen in Fig. \ref{fig5_InverseQualityFactor}(b). 

It is interesting to note that the damping rate of phonons in superfluid
$^{4}$He at $k=0.4\textrm{ }\textrm{Å}^{-1}$ (circles in Fig. \ref{fig5_InverseQualityFactor}(a))
\cite{Stirling1990,DampingRateHe4Note} closely follows our prediction
at small momentum. This similarity between superfluid $^{4}$He and
a unitary Fermi gas suggests that any strongly interacting quantum
fluids may share a \emph{universal} damping rate for phonons, independent
of their entirely different internal structures and quantum statistics.

\textit{Conclusions}. \textemdash{} In summary, we have developed
a finite-temperature microscopic theory of the density response of
a unitary Fermi gas at small transferred momentum, and have quantitatively
examined its reliability by comparing our theoretical results with
the latest Bragg scattering measurements \cite{Hoinka2017}. We have
clarified that the damping rate of the Goldstone phonon mode is largely
due to the inelastic scatterings from fermionic quasiparticles and
have predicted its universal temperature dependence, which is to be
confronted with future experimental confirmation.
\begin{acknowledgments}
We are grateful to Sascha Hoinka and Chris Vale for sharing their
experimental data. Our research was supported by the National Natural
Science Foundation of China, Grant No. 11747059 (PZ), and Australian
Research Council's (ARC) Discovery Projects: FT130100815 and DP170104008
(HH), and FT140100003 and DP180102018 (XJL).
\end{acknowledgments}


\begin{thebibliography}{10}
\bibitem{Nozieres1990}P. Nozières and D. Pines, \textit{Theory of
Quantum Liquids} vol II: \textit{Superfluid Bose Liquids} (Addison-Wesley,
Redwood City, CA, 1990).

\bibitem{Griffin1993}A. Griffin, \textit{Excitations in a Bose-Condensed
Liquid} (Cambridge University Press, New York, 1993).

\bibitem{Khalatnikov1965}I. M. Khalatnikov, \textit{Introduction
to the Theory of Superfluidity} (Benjamin, New York, 1965).

\bibitem{Woods1973}A. D. B. Woods and R. A. Cowley, Rep. Prog. Phys.
\textbf{36}, 1135 (1973).

\bibitem{OHara2002}K. M. O\textquoteright Hara, S. L. Hemmer, M.
E. Gehm, S. R. Granade, and J. E. Thomas, Science \textbf{298}, 2179
(2002).

\bibitem{Randeria2014}M. Randeria and E. Taylor, Annu. Rev. Condens.
Matter Phys. \textbf{5}, 209 (2014).

\bibitem{Bloch2008}I. Bloch, J. Dalibard, and W. Zwerger, Rev. Mod.
Phys. \textbf{80}, 885 (2008).

\bibitem{Regal2004}C. A. Regal, M. Greiner, and D. S. Jin, Phys.
Rev. Lett. \textbf{92}, 040403 (2004).

\bibitem{Navon2010}N. Navon, S. Nascimbène, F. Chevy, and C. Salomon,
Science \textbf{328}, 729 (2010).

\bibitem{Ho2004}T.-L. Ho, Phys. Rev. Lett. \textbf{92}, 090402 (2004).

\bibitem{Hu2007}H. Hu, P. D. Drummond, and X.-J. Liu, Nat. Phys.
\textbf{3}, 469 (2007).

\bibitem{Nascimbene2010}S. Nascimbène, N. Navon, K. J. Jiang, F.
Chevy, and C. Salomon, Nature (London) \textbf{463}, 1057 (2010).

\bibitem{Ku2012}M. J. Ku, A. T. Sommer, L. W. Cheuk, and M. W. Zwierlein,
Science \textbf{335}, 563 (2012).

\bibitem{Sidorenkov2013}L. A. Sidorenkov, M. K. Tey, R. Grimm, Y.-H.
Hou, L. Pitaevskii, and S. Stringari, Nature (London) \textbf{498},
78 (2013).

\bibitem{Kinast2004}J. Kinast, S. L. Hemmer, M. E. Gehm, A. Turlapov,
and J. E. Thomas, Phys. Rev. Lett. \textbf{92}, 150402 (2004).

\bibitem{Bartenstein2004}M. Bartenstein, A. Altmeyer, S. Riedl, S.
Jochim, C. Chin, J. Hecker Denschlag, and R. Grimm, Phys. Rev. Lett.
\textbf{92}, 203201 (2004).

\bibitem{Hu2004}H. Hu, A. Minguzzi, X.-J. Liu, and M. P. Tosi, Phys.
Rev. Lett. \textbf{93}, 190403 (2004).

\bibitem{Altmeyer2007}A. Altmeyer, S. Riedl, C. Kohstall, M. J. Wright,
R. Geursen, M. Bartenstein, C. Chin, J. Hecker Denschlag, and R. Grimm,
Phys. Rev. Lett. \textbf{98}, 040401 (2007).

\bibitem{Hoinka2017}S. Hoinka, P. Dyke, M. G. Lingham, J. J. Kinnunen,
G. M. Bruun, and C. J. Vale, Nat. Phys. \textbf{13}, 943 (2017).

\bibitem{Minguzzi2001}A. Minguzzi, G. Ferrari, and Y. Castin, Eur.
Phys. J. D \textbf{17}, 49 (2001).

\bibitem{Combescot2006}R. Combescot, M. Y. Kagan, and S. Stringari,
Phys. Rev. A \textbf{74}, 042717 (2006).

\bibitem{Stringari2009}S. Stringari, Phys. Rev. Lett. \textbf{102},
110406 (2009).

\bibitem{Zou2010}P. Zou, E. D. Kuhnle, C. J. Vale, and H. Hu, Phys.
Rev. A \textbf{82}, 061605(R) (2010).

\bibitem{Guo2010}H. Guo, C.-C. Chien, and K. Levin, Phys. Rev. Lett.
\textbf{105}, 120401 (2010).

\bibitem{Palestini2012}F. Palestini, P. Pieri, and G. C. Strinati,
Phys. Rev. Lett. \textbf{108}, 080401 (2012).

\bibitem{Hu2012}H. Hu and X.-J. Liu, Phys. Rev. A \textbf{85}, 023612
(2012). 

\bibitem{He2016}L. He, Ann. Phys. (N.Y.) \textbf{373}, 470 (2016).

\bibitem{Zou2016}P. Zou, F. Dalfovo, R. Sharma, X.-J. Liu, and H.
Hu, New J. Phys. \textbf{18}, 113044 (2016).

\bibitem{Vitali2017}E. Vitali, H. Shi, M. Qin, and S. Zhang, Phys.
Rev. A \textbf{96}, 061601(R) (2017).

\bibitem{Kurkjian2017AOP}H. Kurkjian, Y. Castin, and A. Sinatra,
Ann. Phys. (Berlin) \textbf{529}, 1600352 (2017).

\bibitem{Zhang2011}Z. Zhang and W. V. Liu, Phys. Rev. A \textbf{83},
023617 (2011).

\bibitem{Castin2017}Y. Castin, A. Sinatra, and H. Kurkjian, arXiv:1707.09774
(2017).

\bibitem{Kurkjian2017}H. Kurkjian and J. Tempere, arXiv:1708.06249
(2017).

\bibitem{Yu2003}Y. Yu and A. Bulgac, Phys. Rev. Lett. \textbf{90},
222501 (2003).

\bibitem{Bulgac2007}A. Bulgac, Phys. Rev. A \textbf{76}, 040502(R)
(2007).

\bibitem{CutOffMomentumNote}In our numerical calculations, the cut-off
momentum $\Lambda$ is sent to $\infty$, as all the final equations
are free from divergence due to renormalization.

\bibitem{EffectiveMassNote}In principle, the effective mass $m^{*}$
of Bogoliubov quasiparticles differs from the bare mass $m$ of fermions.
However, this difference is very small near the unitary limit \cite{Yu2003,Bulgac2007},
so we set $m^{*}=m$ for simplicity. This simple choice also ensures
that the $f$-sum rule of the dynamic structure factor is strictly
satisfied.

\bibitem{Carlson2014}J. Carlson and S. Gandolfi, Phys. Rev. A \textbf{90},
011601(R) (2014).

\bibitem{Brunello2001}A. Brunello, F. Dalfovo, L. Pitaevskii, S.
Stringari, and F. Zambelli, Phys. Rev. A \textbf{64}, 063614 (2001).

\bibitem{Blakie2002}P. B. Blakie, R. J. Ballagh, and C. W. Gardiner,
Phys. Rev. A \textbf{65}, 033602 (2002).

\bibitem{SpectrumConvolutionNote}We note that, if we swap the order
of two convolutions, the resulting spectrum remains essentially the
same. The convolution operation also does not violate the $f$-sum
rule for the density response function and dynamic structure factor. 

\bibitem{Hu2006}H. Hu, X.-J. Liu, and P. D. Drummond, Europhys. Lett.
\textbf{74}, 574 (2006). The predicted equations of state by this
Gaussian pair fluctuation theory agree excellently well with the measured
equations of state at BEC-BCS crossover at nearly zero temperature,
see, for example, Fig. 3(a) in Ref. \cite{Navon2010}. 

\bibitem{Kurkjian2016}H. Kurkjian, Y. Castin, and A. Sinatra, Phys.
Rev. A \textbf{93}, 013623 (2016).

\bibitem{Haussmann2009}R. Haussmann, M. Punk, and W. Zwerger, Phys.
Rev. A \textbf{80}, 063612 (2009).

\bibitem{ZhangLiuResultNote}The analytic expression derived by Zhang
and Liu takes the form \cite{Zhang2011}, $\Gamma/\omega_{0}=(3\pi/2)(c_{s}/v_{F})^{3}(1+\Delta^{2}/\xi_{p}^{2})^{2}e^{-E(p)/k_{B}T}$,
where the wavevector $p$ satisfies $\hbar p\xi_{p}=mc_{s}E_{p}$
with $\xi_{p}\equiv\hbar^{2}p^{2}/(2m)-\mu$ and $E_{p}=\sqrt{\xi_{p}^{2}+\Delta^{2}}$.
By setting $\mu=0.376\varepsilon_{F}$ and $\Delta=0.47\varepsilon_{F}$,
we find that $p\simeq0.784k_{F}$, $E(p)\simeq0.527\varepsilon_{F}$,
and $\Gamma/\omega_{0}\simeq5.03e^{-0.527T_{F}/T}$. The last expression
gives the blue dot-dashed line in Fig. \ref{fig5_InverseQualityFactor}(a).

\bibitem{Stirling1990}W. G. Stirling and H. R. Glyde, Phys. Rev.
B \textbf{41}, 4224 (1990).

\bibitem{DampingRateHe4Note}If we treat the wavevector of rotons
at saturated vapor pressure, $k_{R}=1.925\textrm{ }\textrm{Å}^{-1}$,
as the Fermi wavevector $k_{F}^{\textrm{He}}$ for $^{4}$He atoms,
then, the case of $k=0.4\textrm{ }\textrm{Å}^{-1}$ corresponds to
$k/k_{F}^{\textrm{He}}\simeq0.2$ and may be viewed as the limit of
small transferred momentum $k\rightarrow0$.
\end{thebibliography}
\end{document}